\documentclass{pj}
\usepackage{graphicx}

\begin{document}
\setcounter{page}{1}
\pjheader{2020} 

\title{Quasi Modes and Density of States (DOS) of 1D Photonics Crystal}

\author{Shaolin Liao$^{1, *}$ and Lu Ou$^2$}

\address{$^1$ (2008-2010) Physics Department, Queens College, City University of New York, 65-30 Kissena Blvd, Flushing NY 11367; \\$^2$ College of Computer Science and Electronic Engineering, Hunan University, Changsha, Hunan, China 410082. \\ $^*$Corresponding author (sliao5@iit.edu): S. Liao is now at the Department of Electrical and Computer Engineering, Illinois Institute of Technology, Chicago, IL, USA 60616.
}

\runningauthor{Liao}
\tocauthor{S.~Liao}

\begin{abstract}
1-Dimensional (1D) photonics crystals  with and without defects have been numerically studied using efficient Transfer Matrix Method (TMM). Detailed numerical recipe of the TMM has been laid out.  Dispersion relation is verified for the periodic Photonics Band Gap (PBG) structure.  When there are defects, the transmission spectrum can be decomposed into one or more quasi modes with excellent agreement. The Density of States (DOS) is obtained from the phase derivative of the transmission spectrum. Green's function is also obtained showing much sharper mode characteristics when the excitation source is localized at the peaks of the quasi modes.
\end{abstract}

%

\section{Introduction}
\label{secttion1}

Photonics crystal  has been of great interest due to its potential applications in optics filter and low-loss reflection mirror \cite{Yablonovich},   \cite{Wu}. The transmission spectrum and its phase information of different kinds of photonics crystal has been widely reported using Transfer Matrix Method (TMM) \cite{Wu}. Experiment on single-mode 1-dimensional (1D) waveguide system has been recently carried out in our group \cite{Sebbah}.  What's more, the Green's function is important in description of random laser \cite{Tureci_PRA}  and periodic Photonics Band Gap (PBG) structure with defects \cite{Sebbah}  since the radiation can be understood as convolution of the Green's function with the source. Also, Local Density of States (LDOS) is closely related to the imaginary part of the Green's function \cite{Sheng}. The Density of States (DOS), which is the volume integral of the LDOS,  is shown to be  the phase derivative of the transmission spectrum with respect to frequency for 1D system without loss \cite{Avishai}.

 In this article, we numerically study the transmission spectrum, its phase information and the Green's function, by means of TMM \cite{Wu}.  The obtained 1D transmission spectrum is further decomposed into quasi modes \cite{Ching}.   The DOS is obtained from the phase derivative of the transmission spectrum.

\section{Transfer Matrix Method}
\label{section2}
The electromagnetic wave is governed by the Maxwell's equations and has many applications \cite{Liao_Ping_Pong_APMC_2020}-\cite{Liao_Polarization_FIO_2008}. Here we are dealing with the 1D problem in the optics regime.
\subsection{Green's function}

The Green's function is described by

 \begin{equation}\label{wave}
 \nabla^2 G(x, x') + k^2(x) G(x, x')    = -\delta (x - x')
\end{equation}
where the delta source is located at $x'$.

The boundary conditions are

\begin{eqnarray}\label{Dirichlet}
 G(x_b^+, x')  &=&  G(x_b^-, x') \\
  \frac{d G(x_b^+, x')}{d x}   &-& \frac{d G(x_b^-, x')}{d x}   = -\delta (x_b - x') \nonumber
\end{eqnarray}
where $x_b^\pm$ denote the right and left sides of  the boundary.

\subsection{Forward- and backward- propagating waves}

The Green's function $G(x, x')$ in each layer can be expressed as superposition of forward- and back-ward propagating waves,

\begin{equation}\label{forward_backward}
  G(x, x') = t  \exp^{- j k_0 (n_r   - jn_i) x} + r  \exp^{j k_0 (n_r - jn_i) x}
\end{equation}
  where $n_r - j n_i$ is the complex index of refractive.

   \subsubsection{On the layer boundary}
   Substituting  Eq. (\ref{forward_backward}) into Eq. (\ref{Dirichlet})   and setting $x_b =0$, we have,
   \begin{eqnarray}\label{matrix1}
   \underline{v}^\pm &=& T^\pm \underline{v}^\mp + S^\pm \underline{u}
   \end{eqnarray}
where the superscript "+" denotes the right side of the boundary and "-" denotes the left side of the boundary. The transfer matrixes $T^\pm$ and source matrixes $S^\pm$ are given below,
\begin{eqnarray}
       T^\pm &=&     \frac{1}{2}\left[\begin{array}{cc} 1 + \frac{n_r^\mp - j n_i^\mp}{n_r^\pm - j n_i^\pm} &   1 - \frac{n_r^\mp - j n_i^\mp}{n_r^\pm - j n_i^\pm} \\ 1 - \frac{n_r^\mp - j n_i^\mp}{n_r^\pm - j n_i^\pm} & 1 + \frac{n_r^\mp - j n_i^\mp}{n_r^\pm - j n_i^\pm}       \end{array} \right] \nonumber \\
       S^\pm &=& \pm \frac{1}{2}\left[\begin{array}{cc} 1  &   \frac{1}{n_r^\pm - j n_i^\pm} \\ 1  &   \frac{-1}{n_r^\pm - j n_i^\pm}       \end{array} \right]   \nonumber
\end{eqnarray}

 \subsubsection{Wave propagations inside layer}

 The wave propagations inside each layer can be described by a propagators $P^\pm$,
   \begin{eqnarray}\label{Pp}
   \underline{v}_{m}^\pm = P_m^\pm   \underline{v}_{m}^\mp
\end{eqnarray}
where $\pm$  denote the left and right ends of layer $m$ respectively, and
 \begin{eqnarray}\label{Ppm}
   P_m^\pm   =  \left[ \begin{array}{cc} \exp^{\mp j k_0 (n_r^m - j n_i^m) d_m} &   0  \\ 0 &  \exp^{ \pm j k_0 (n_r^m - j n_i^m)d_m}     \end{array} \right] \nonumber
\end{eqnarray}
 where $d_m$ is the thickness of layer $m$.

 \subsubsection{The excitation source}

  From Eq. (\ref{matrix1}), we know that, away from the source location $x'$,

    \begin{eqnarray}\label{Transfer}
   \underline{v}^\pm &=& T^\pm \underline{v}^\mp
   \end{eqnarray}

 At the excitation source location, we have

  \begin{eqnarray}\label{source}
   \underline{v}^+ &=&  \underline{v}^- +\left[\begin{array}{cc} \frac{-j}{2k_0} \frac{1}{n_r  - j n_i }  \\  \frac{j}{2k_0} \frac{1}{n_r  - j n_i } \end{array} \right]
   \end{eqnarray}
where $n_r - j n_i$ is complex index of refractive at the source location $x'$.

\begin{figure}[h]
 \centerline{\includegraphics[width=1\textwidth]{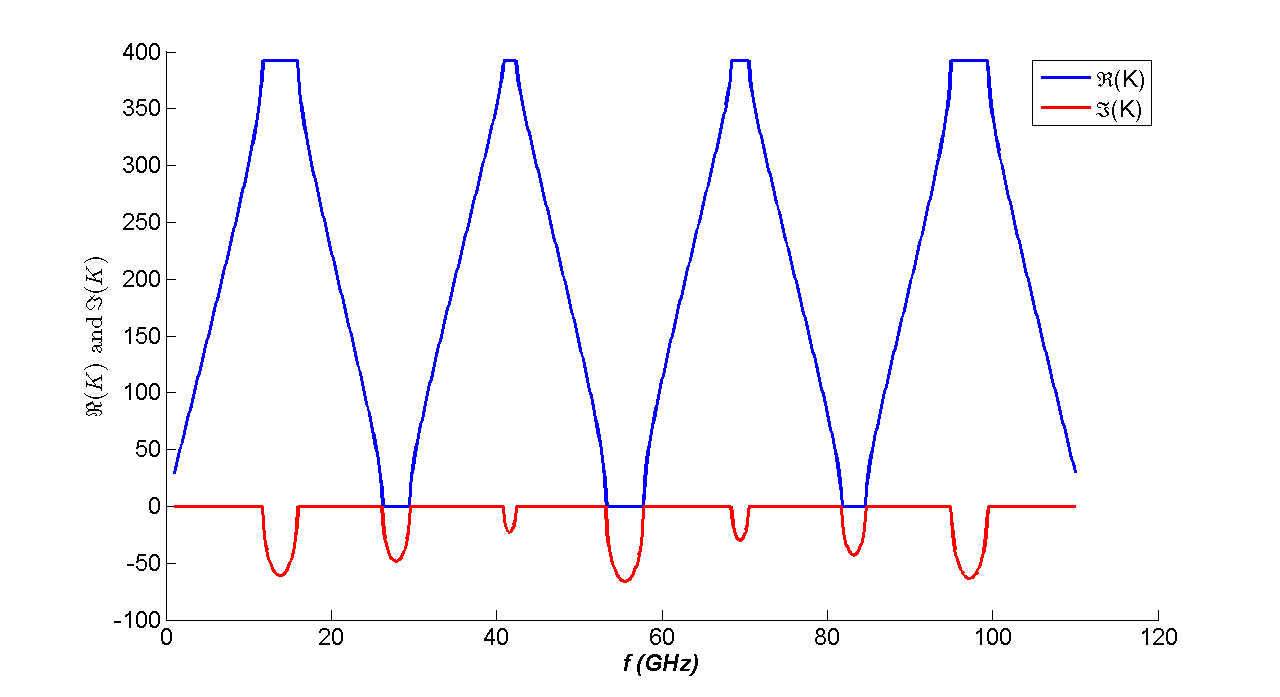} }
\caption{Dispersion relation of a typical PBG structure: $a = 2 d_1 = 2 d_2 = 8$ mm; $n_1 = 1$ and $n_2 = 1.7$.}
\label{kronig}
\end{figure}

 \subsection{Numerical recipe}\label{recipe}

 The numerical recipe to obtain the 1D Green's function is as follows, 1) individually, calculate the transmission and reflection spectra for the segments to the left and to the right sides of the source location $x'$, through Eq. (\ref{Pp}) and Eq. (\ref{Transfer}); 2) connect both segments at the source location, through Eq. (\ref{source}).

\subsubsection{The left and right segments}

For each segment, we have to cascade the propagator $P^-$ in  Eq. (\ref{Pp}) and the transfer matrix $T^-$ in Eq. (\ref{Transfer}), i.e.,  the cascading procedure has to been done from the end of each segment, where there is only forward propagating wave,

 \begin{eqnarray}\label{cascade}
   \underline{v}^+ (\hbox{left/right}) = \left[\prod_{m=1}^N  P_m^-  T_m^-  \right] \left[ \begin{array}{cc} 1  \\ 0  \end{array} \right]
\end{eqnarray}

\subsubsection{Connect both segments}

We now connect both segments at the source location through Eq. (\ref{source}),

  \begin{eqnarray}\label{tsource}
   t(\hbox{right}) &=& t(\hbox{left})  -  \frac{j}{2k_0} \frac{1}{n_r  - j n_i} \\
    r(\hbox{right}) &=& r(\hbox{left})  +  \frac{j}{2k_0} \frac{1}{n_r - j n_i} \nonumber
   \end{eqnarray}

\section{Periodic PBG Structure}
\label{section3}

\begin{figure}[h]
 \centerline{\includegraphics[width=1\textwidth]{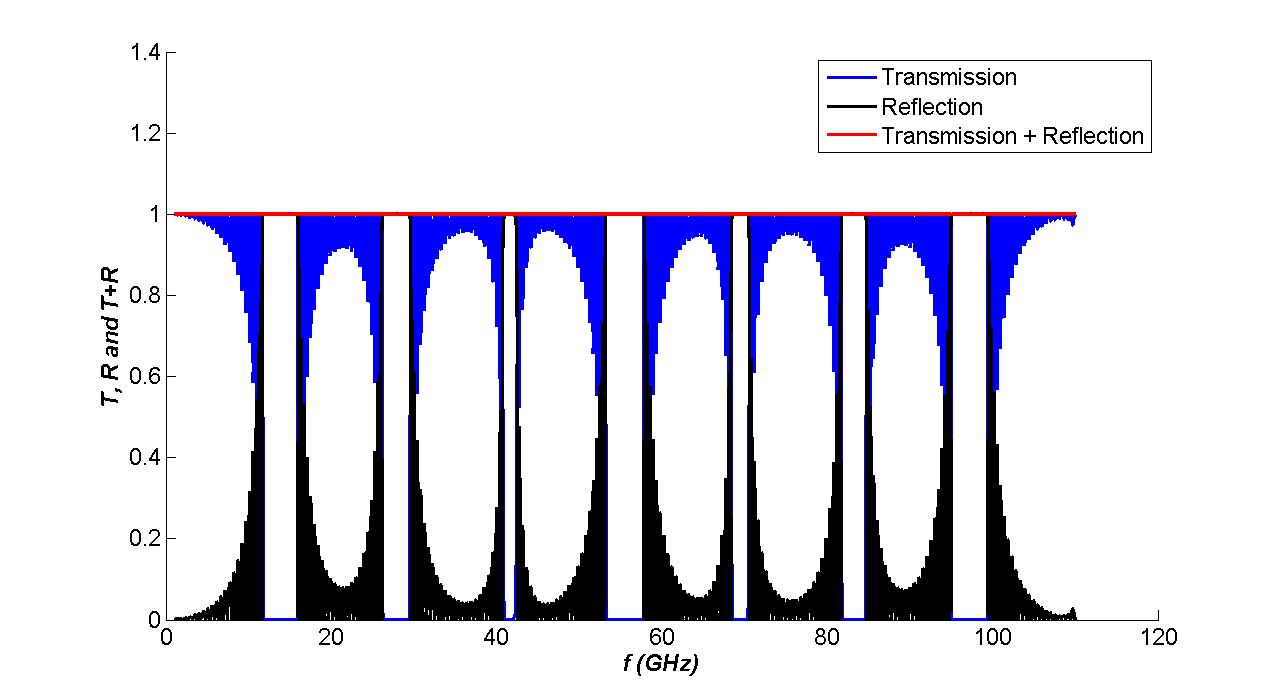} }
\caption{Transmission intensity $T$, reflection intensity $R$ and their sum $T + R$ for 25 pairs of dielectric layers.}
\label{tr}
\end{figure}

 It is well-known that  the dispersion relation of the periodic binary dielectric layers is given by \cite{Stancil},

    \begin{eqnarray}\label{dispersion}
    \cos(K a) = \cos\left(k_1 d_1\right) \cos\left(k_2 d_2\right) - \frac{k_1^2 + k_2^2}{2 k_1 k_2}\sin\left(k_1 d_1\right) \sin\left(k_2 d_2\right)
   \end{eqnarray}
where $K$ is the wave vector of the Bloch wave; $k_{1,2} = \omega \sqrt{\mu \epsilon_{1,2}}$ are the wave vector of the binary pair; $a = d_1 + d_2$ is the periodicity of the structure, with $d_{1,2}$ being the thickness of each layer. Fig. \ref{kronig} shows a typical dispersion relation for $a = 2 d_1 = 2 d_2 = 8$ mm; $n_1 = 1$ and $n_2 = 1.7$; the frequency ranges from 1 GHz  to 110 GHz. The simulated transmission spectrum for such periodic PBG structure has been carried out for 25 pairs of binary dielectric layers: incident wave from left and transmitted wave on the right. The result is shown in Fig. \ref{tr},  which agrees well with Fig. \ref{kronig}.

\begin{figure}[h]
 \centerline{\includegraphics[width=1\textwidth]{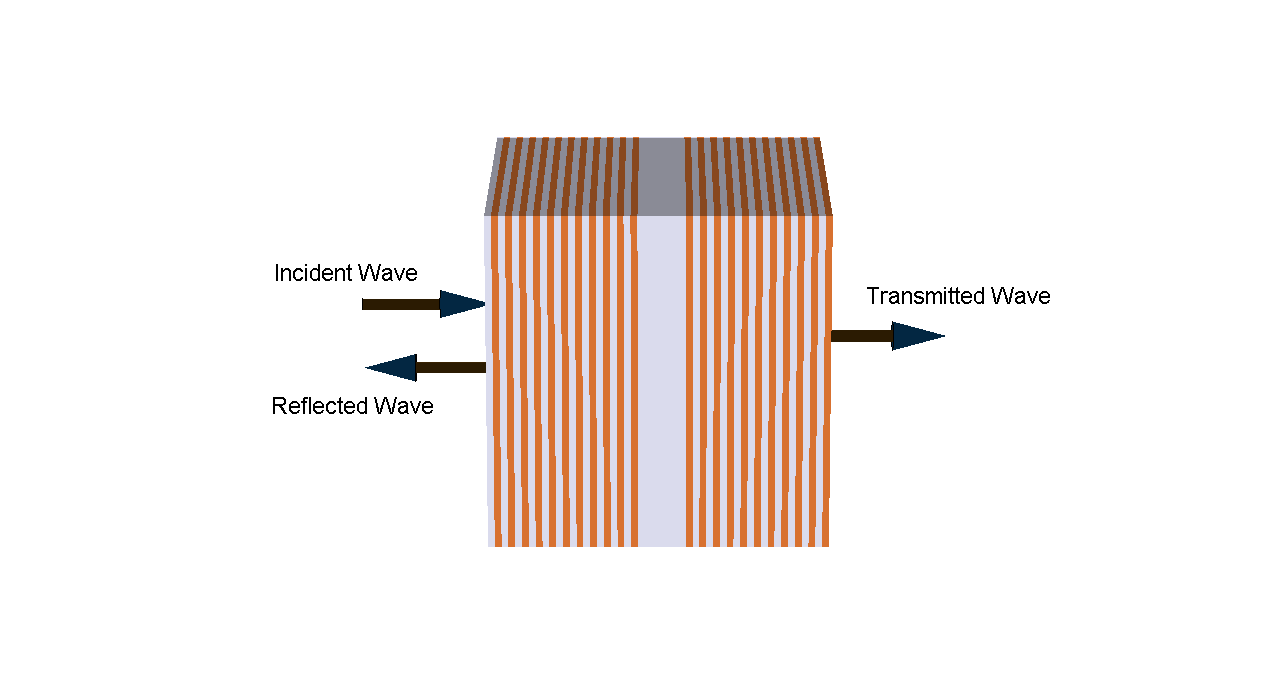} }
\caption{Single defect inside PBG structure is simulated: 3 pairs in the middle are replaced with $n_1$. Gray: $n_1 = 1$; Orange: $n_2 = 1.7$.}
\label{layer_defect}
\end{figure}

\begin{figure}[h]
 \centerline{\includegraphics[width=1\textwidth]{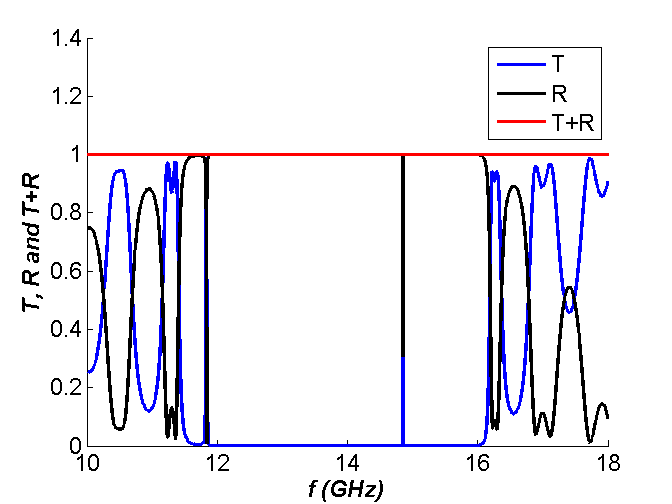} }
\caption{First band gap of PBG structure with single defect: transmission intensity $T$, reflection intensity $R$ and their sum $T + R$.}
\label{single_defect}
\end{figure}

\begin{figure}[h]
 \centerline{\includegraphics[width=1\textwidth]{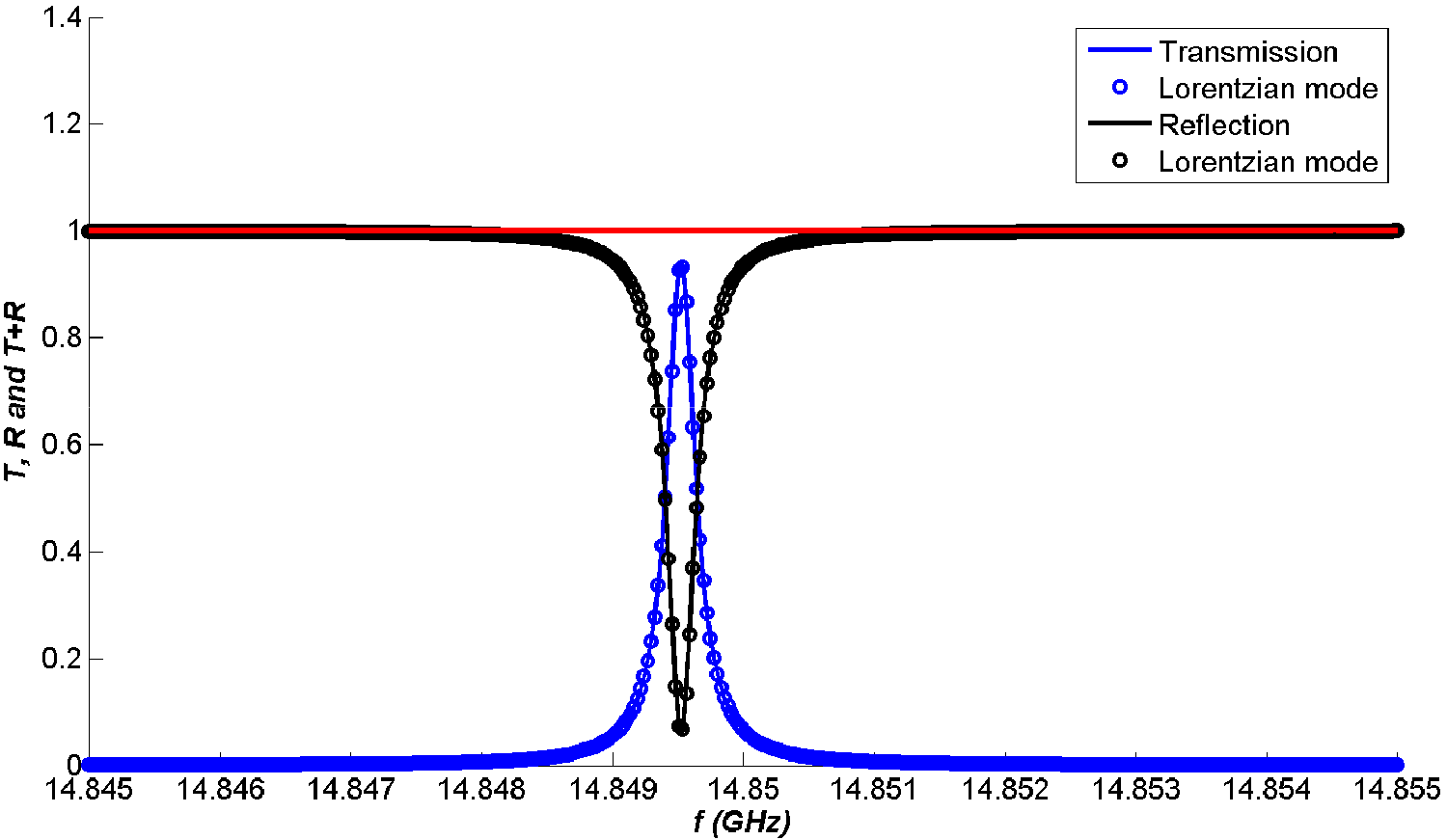} }
\caption{Single defect (a closer look of Fig. \ref{single_defect}): simulated intensity and that of the theoretical Lorentzian quasi mode, with $f_c = 14.85$ GHz and $\Gamma = 0.131$ MHz.}
\label{QNM}
\end{figure}
\begin{figure}[h]
 \centerline{\includegraphics[width=1\textwidth]{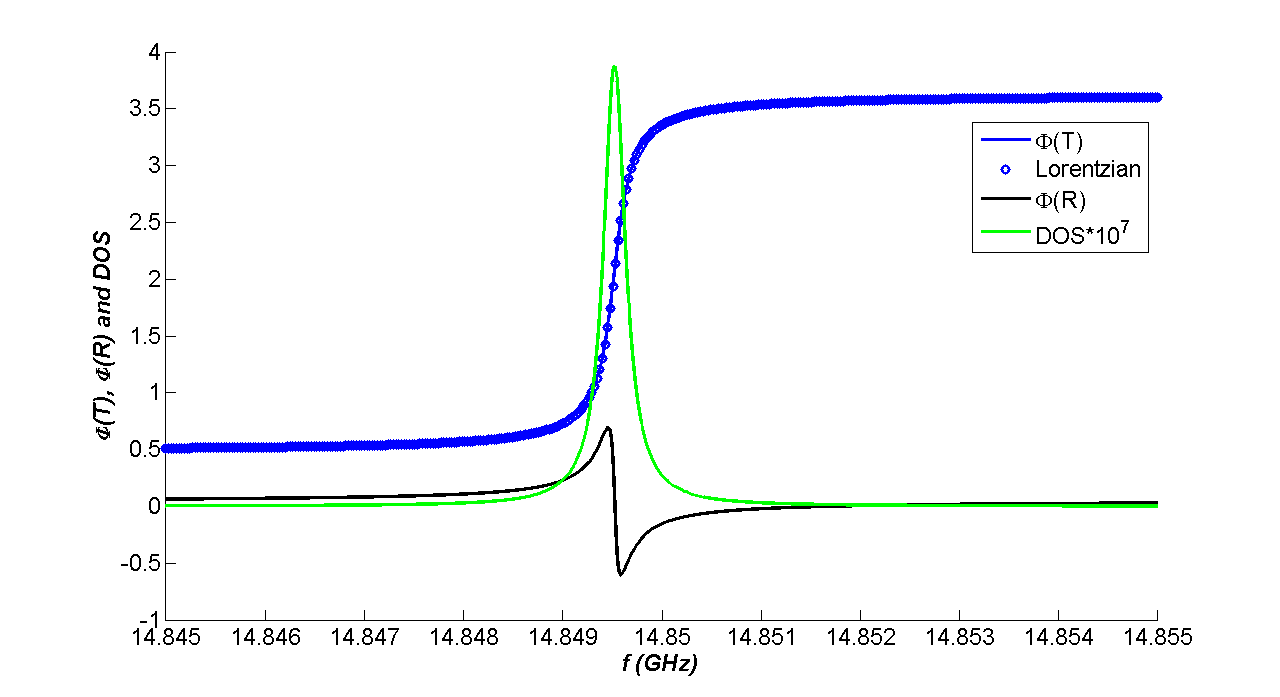} }
\caption{Single defect: simulated phase and and that of the Lorentzian quasi mode, with $f_c = 14.85$ GHz and $\Gamma = 0.131$ MHz.}
\label{QNM_phase}
\end{figure}

 \section{Defects and Quasi Modes}
\label{section4}

\subsection{Transmission and reflection spectra}
Now let's look at what happens when one or more defects are introduced deep inside the PBG structure (25 pairs). First, let's look at single potential well (defect) by replacing the middle 3 pairs with $n_1$ (see Fig. \ref{layer_defect}). The result is shown in Fig. \ref{single_defect} for the first band gap in Fig. \ref{tr}.  One quasi modes appears at around $f_c = 14.85$ GHz.  A closer look at the intensity and phase of the transmission and reflection spectrum are shown in  Fig. \ref{QNM} and Fig. \ref{QNM_phase}, together with the quasi mode fitting and DOS. Quasi modes will be explained in Section \ref{quasimode} and DOS is discussed in Section \ref{sec:dos}.

\begin{figure}[h]
 \centerline{\includegraphics[width=1\textwidth]{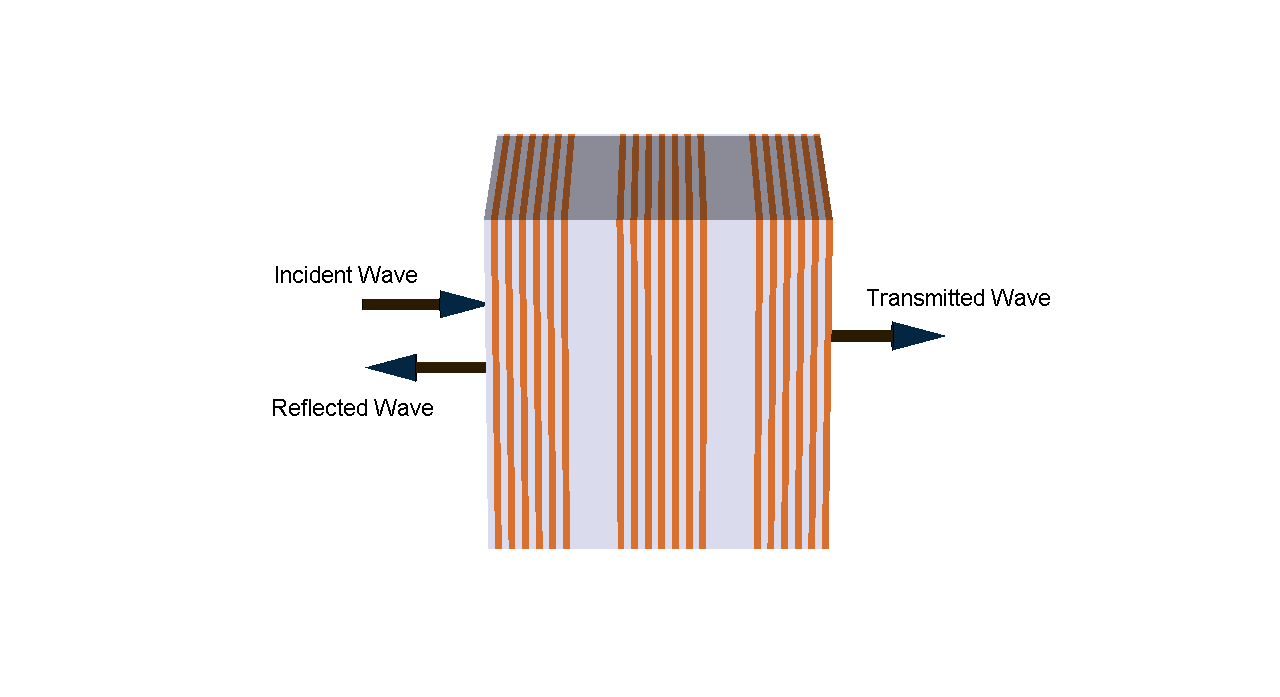} }
\caption{Double defects inside PBG structure is simulated: 3 pairs at two locations  are replaced with $n_1$, separating by a potential barrier $n_2$. Gray: $n_1 = 1$; Orange: $n_2 = 1.7$.}
\label{layer_defect2}
\end{figure}

\begin{figure}[h]
 \centerline{\includegraphics[width=1\textwidth]{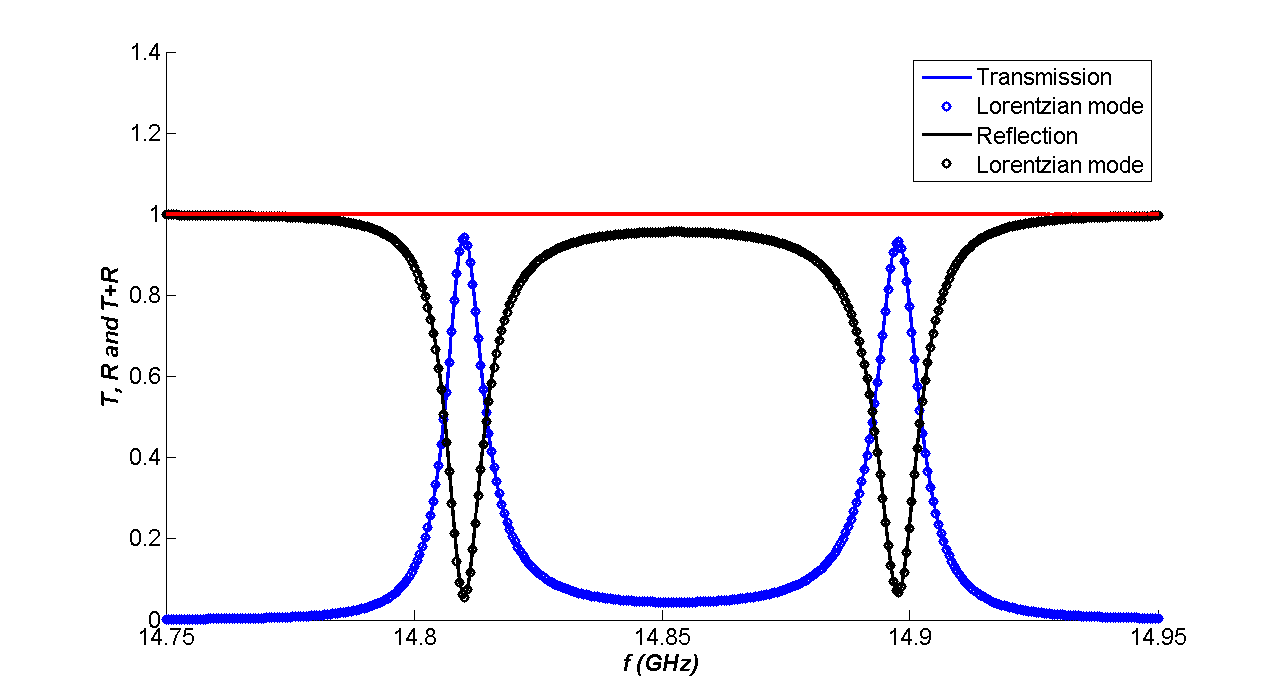} }
\caption{Double defects: simulated transmission and reflection intensities and those of the sum of two Lorentzian quasi modes, with $f_1 = 14.810$ GHz,  $\Gamma_1 = 4.5$ MHz and $f_2 =  14.898$ GHz,  $\Gamma_2 = 5.1$ MHz.}
\label{Defect2}
\end{figure}

\begin{figure}[h]
 \centerline{\includegraphics[width=1\textwidth]{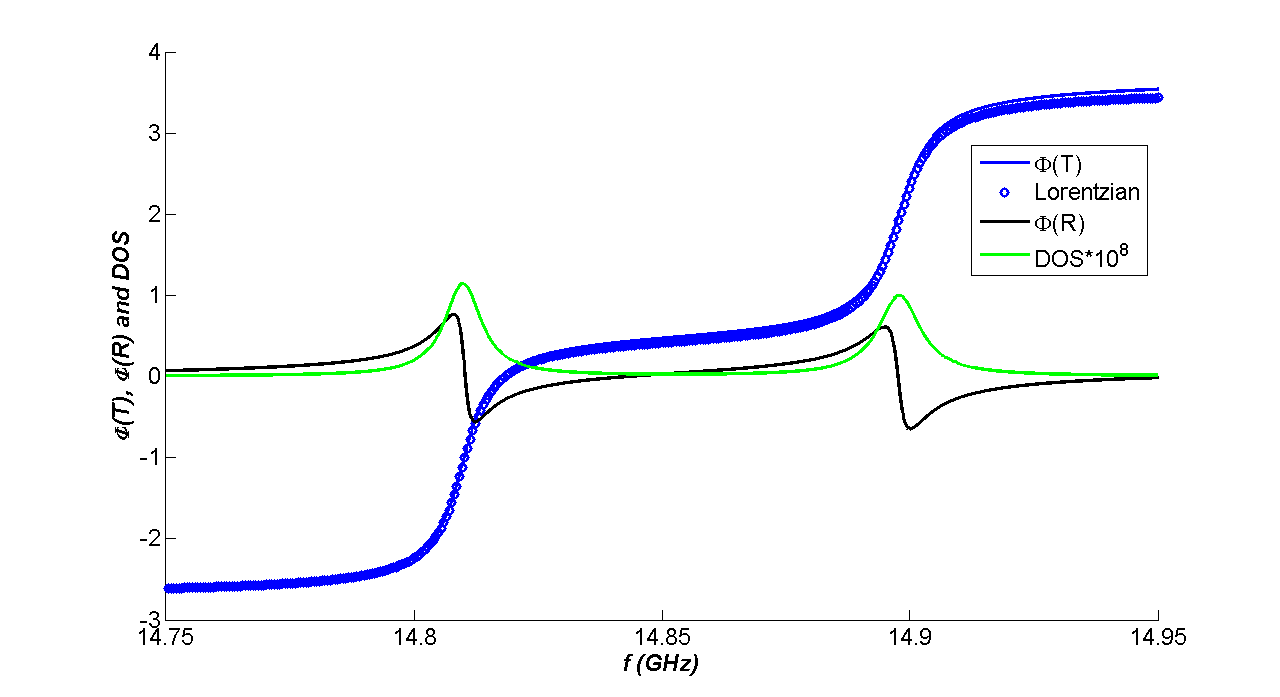} }
\caption{Double defects: simulated phase and and that  of the sum of two Lorentzian quasi modes, with $f_1 = 14.810$ GHz,  $\Gamma_1 = 4.5$ MHz and $f_2 =  14.898$ GHz,  $\Gamma_2 = 5.1$ MHz.}
\label{Defect2_phase}
\end{figure}

Now let's look at double potential wells (defects) by replacing 3 pairs with $n_1$ (see Fig. \ref{layer_defect2}) at two locations separated by a potential barrier $n_2$. The intensity and phase are shown in Fig. \ref{Defect2} and Fig. \ref{Defect2_phase} respectively. Also see Section \ref{quasimode} for quasi mode explanation and Section \ref{sec:dos} for DOS.

 $ $


 \subsection{Quasi modes}\label{quasimode}

Quasi modes can be considered as localized modes  around the defect sites and could couple with each other if more than one  modes are present. Mathematically, the frequency part of the quasi mode can be expressed as the Lorentzian function,

   \begin{eqnarray}\label{Lorentzian}
    \Psi(f)  = \frac{\Gamma}{ (f - f_c) + j \Gamma}
   \end{eqnarray}

\subsubsection{Single quasi mode}

In Fig. \ref{QNM} and Fig. \ref{QNM_phase}, we look closer into the quasi mode shown in Fig. \ref{single_defect}. We also plot the Lorentzian quasi mode as circles in Fig. \ref{QNM} and Fig. \ref{QNM_phase}, which agrees with the simulation very well. We found that $f_c = 14.85$ GHz and $\Gamma = 0.131$ MHz for this quasi mode.

\subsubsection{Two quasi modes}

For double defects inside PBG structure., the wave function can be expressed in sum of two quasi modes,

   \begin{eqnarray}\label{Lorentzian2}
    \Psi(f)  = a_1 \frac{\Gamma_1}{ (f - f_1) + j \Gamma_1} + a_2 \exp^{j \pi} \frac{\Gamma_2}{ (f - f_2) + j \Gamma_2}
   \end{eqnarray}
  with $f_1 = 14.810$ GHz,  $\Gamma_1 = 4.5$ MHz, $a_1 = 0.972$ and $f_2 =  14.898$ GHz,  $\Gamma_2 = 5.1$ MHz, $a_2 = 0.966$. Note that in Eq. (\ref{Lorentzian2}), the phase difference between these adjacent two modes is $\pi$.

   \subsection{Density of States} \label{sec:dos}

  For 1D system, DOS is given by \cite{Avishai},
        \begin{eqnarray}\label{DOS}
        \hbox{DOS}  \ \ = \frac{1}{\pi} \frac{d \phi}{d \omega}
   \end{eqnarray}
   We also plot the DOS for single defect and double defects in Fig. \ref{QNM_phase} and Fig. \ref{Defect2_phase} respectively.

\begin{figure}[h]
 \centerline{\includegraphics[width=1\textwidth]{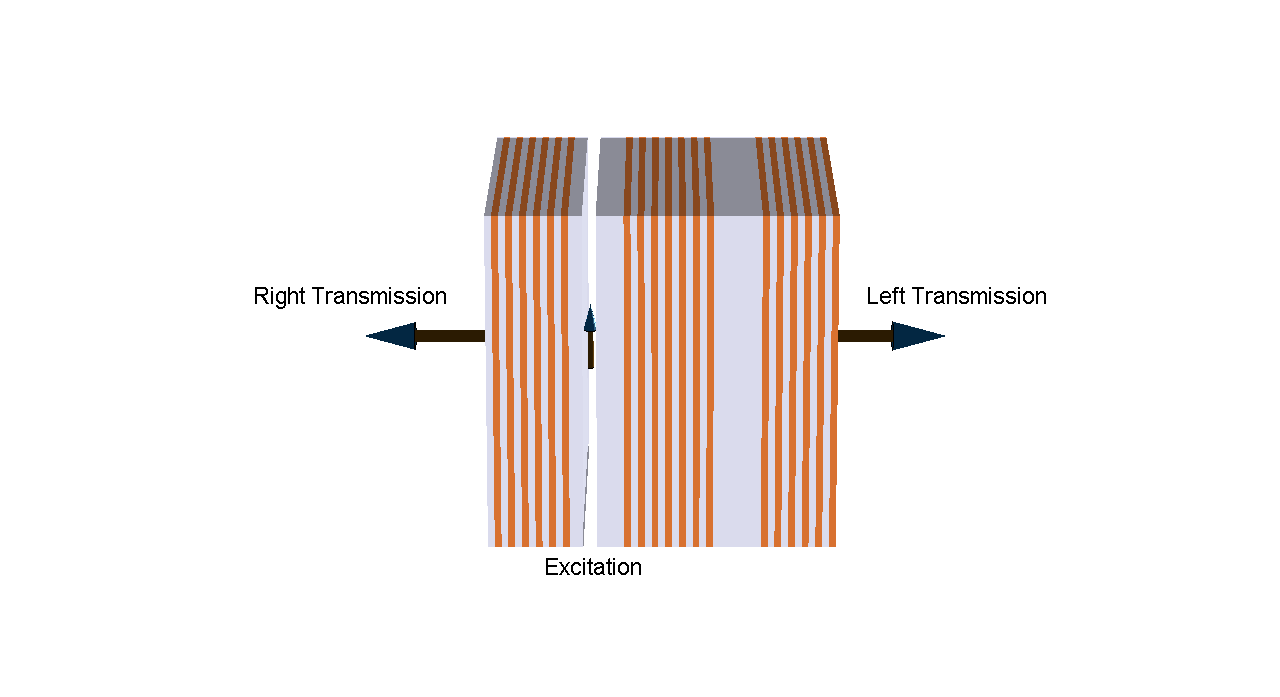} }
\caption{Schematics of Green's function for double defects. Gray: $n_1 = 1$; Orange: $n_2 = 1.7$.}
\label{layer_defect2_green}
\end{figure}

\begin{figure}[h]
 \centerline{\includegraphics[width=1\textwidth]{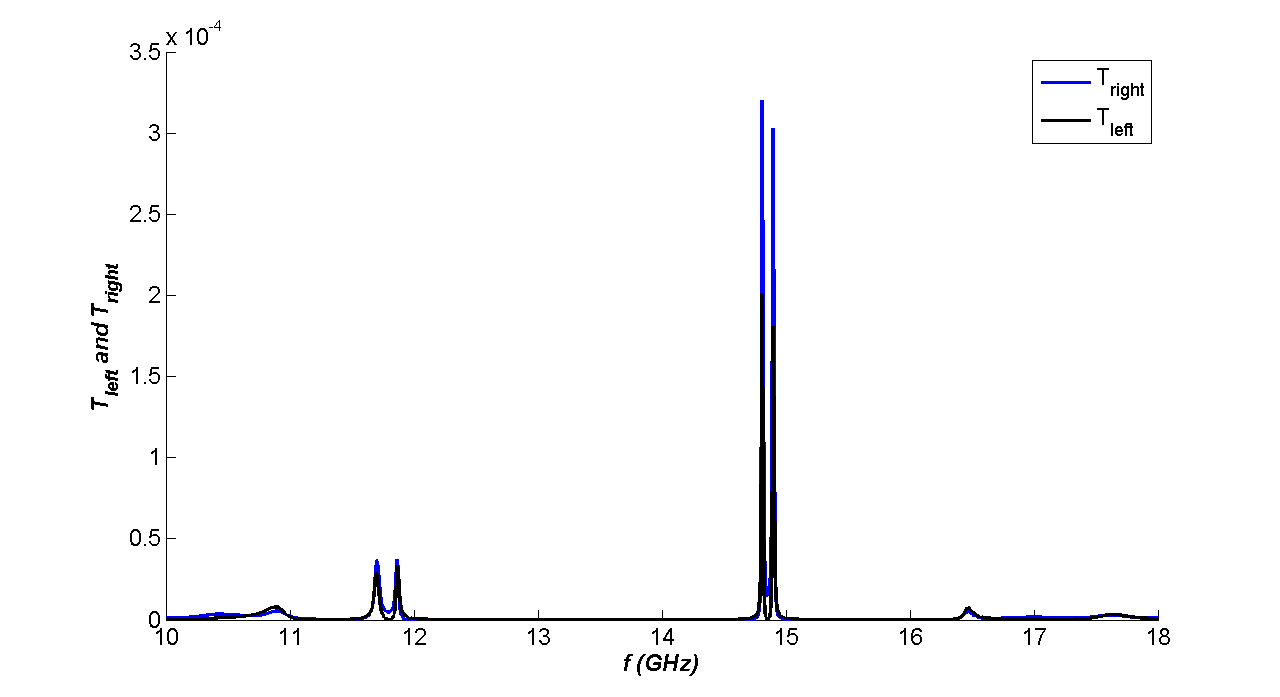} }
\caption{Simulated transmission spectrum of Green's function for both left and right transmissions.}
\label{Defect2_green}
\end{figure}
\begin{figure}[h]
 \centerline{\includegraphics[width=2.5 in, height= 1.6 in]{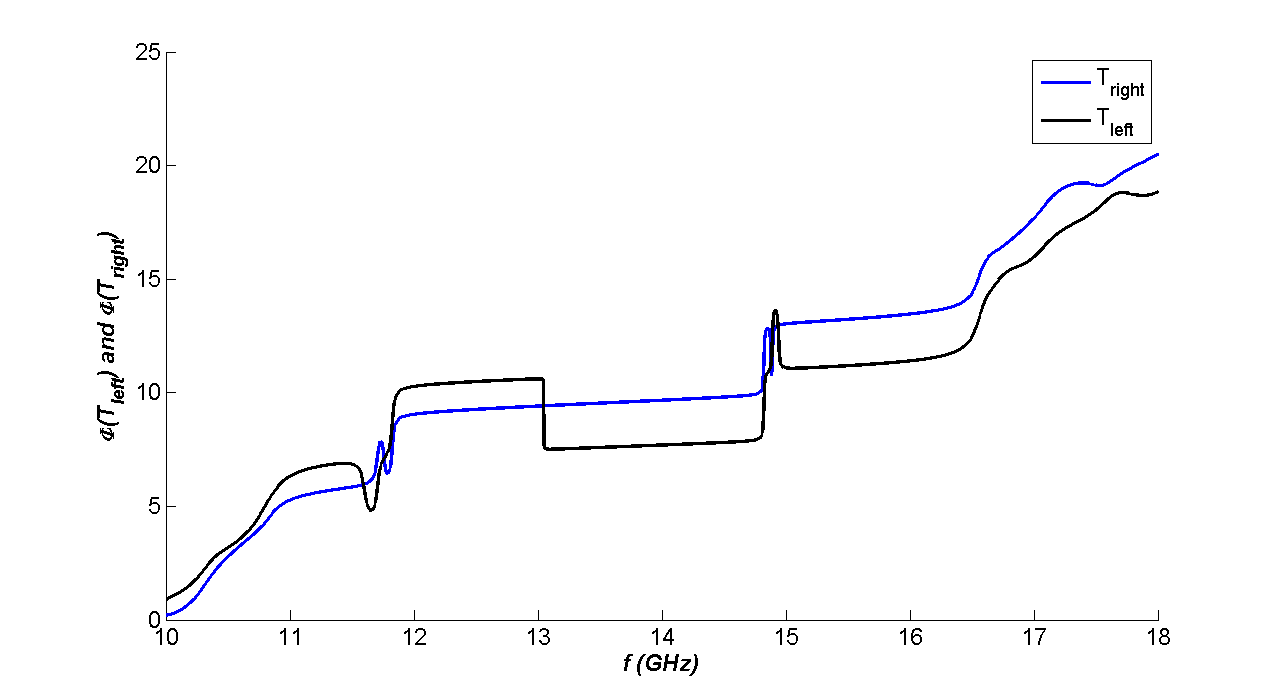} }
\caption{Simulated phase of the transmission spectrum of Green's function for both left and right transmissions.}
\label{Defect2_green_phase}
\end{figure}

\subsection{Green's function}

We also obtained  the Green's function following the procedure stated in Section \ref{recipe}. Here we show the result  for double defects: the schematics with excitation source inside the first defect is shown in Fig. \ref{layer_defect2_green}. The transmission spectra and phases for both left and right sides are shown in Fig. \ref{Defect2_green} and Fig. \ref{Defect2_green_phase} respectively. Compared to the transmission and reflection spectra in Fig. \ref{Defect2} and Fig. \ref{Defect2_phase}, we can see that the Green's function shows much sharper quasi mode peaks.

\section{Conclusion}
We have simulated the transmission spectrum and the Green's function for 1D photonics crystal with and without defects. It has been shown that quasi mode decomposition gives excellent agreement with the simulated result.   DOS is also obtained through the phase derivative of the transmission spectrum.  The method can find applications in  design of 1D photonics filter and reflection mirror, defect simulation of photonics crystal and its DOS.

\ack
Shaolin Liao wants to thank Prof. Azriel Genack and his group for their help when he worked there as a Postdoc Fellow from 2008-2010.

\end{document}